\documentclass[letterpaper]{article}
\usepackage[preprint]{aaai2027}

\usepackage{courier}  
\usepackage[hyphens]{url}  
\usepackage{graphicx} 
\usepackage{natbib}  
\usepackage{caption} 
\usepackage{algorithm}
\usepackage{algorithmic}
\usepackage{enumitem}
\usepackage{xcolor}
\usepackage{amsmath}
\usepackage{amssymb}

\usepackage{xcite}
\usepackage{amsfonts}
\usepackage{textcomp}

\usepackage{multirow}
\usepackage{bigstrut}
\usepackage{makecell}
\usepackage{pifont}
\usepackage{subfloat}
\usepackage{diagbox} 
\usepackage{placeins}
\usepackage{boldline}
\usepackage{booktabs}
\usepackage{subcaption}
\usepackage{framed}
\usepackage{color}
\usepackage{array}
\definecolor{shadecolor}{gray}{1} 

\usepackage[inkscapelatex=false]{svg}

\usepackage{cleveref}
\crefname{proposition}{Proposition}{Propositions}

\usepackage{newfloat}
\usepackage{listings}
\DeclareCaptionStyle{ruled}{labelfont=normalfont,labelsep=colon,strut=off} 
\floatstyle{ruled}
\newfloat{listing}{tb}{lst}{}
\floatname{listing}{Listing}
\title{DDSNet: Dual-domain Symmetry-aware Network for PCSEL Property Prediction}
\author{
Cen Chen\equalcontrib\textsuperscript{\rm 1,\rm 2},
Haitao Huang\equalcontrib\textsuperscript{\rm 1},
Jiazhi Mao\textsuperscript{\rm 3},
Feifan Xu\textsuperscript{\rm 3},
Zhe Zhuang\corresponding\textsuperscript{\rm 3},
Yuxiang Ren\corresponding\textsuperscript{\rm 1}
}

\affiliations{
\textsuperscript{\rm 1}School of Intelligent Science and Technology, Nanjing University\\
\textsuperscript{\rm 2}School of Information and Software Engineering, University of Electronic Science and Technology of China\\
\textsuperscript{\rm 3}School of Integrated Circuits, Nanjing University\\
ccen1613@gmail.com,
casishaitaohuang@gmail.com,
jzmao@smail.nju.edu.cn,
ffxu@nju.edu.cn,
zzhuang@nju.edu.cn,
renyuxiang@nju.edu.cn
}

\usepackage{bibentry}

\begin{document}
\maketitle

%


\begin{abstract}
Efficient exploration of the photonic crystal (PhC) lattice design space is essential for developing photonic crystal surface-emitting lasers. While coupled-wave theory (CWT) provides an effective physical framework, its computational cost remains prohibitive for large-scale exploration, driving the demand for neural surrogates. However, existing AI models underexploit two key factors of PhC unit-cell dielectric patterns indicated by CWT: spectral components and asymmetric structures, which largely govern devices' physical properties. This mismatch weakens surrogate accuracy and screening reliability, especially in structure-sensitive regions. To address this, we propose the Dual-Domain Symmetry-Aware Network (DDSNet). It integrates translation-equivariant spectral filtering with a symmetry-induced structural prior. The spectral filtering injects a spectral inductive bias into vision model while preserving translation equivariance on lattices. Meanwhile, the structural prior decomposes lattice features into irreducible representation-associated, symmetry-resolved components and processes them in separate branches. Experiments demonstrate that DDSNet significantly outperforms existing AI baselines in property prediction and high-throughput screening, exhibiting superior reliability in structure-sensitive regions. Crucially, component masking analyses reveal that the network successfully learns property-specific dependencies aligned with physical priors. These results indicate that DDSNet effectively captures physically meaningful structure-property relationships, establishing a highly reliable neural surrogate for PhC design space exploration.
\end{abstract}

\section{Introduction}
\label{introduction}

\begin{figure}
    \centering
    \includegraphics[width=1\linewidth]{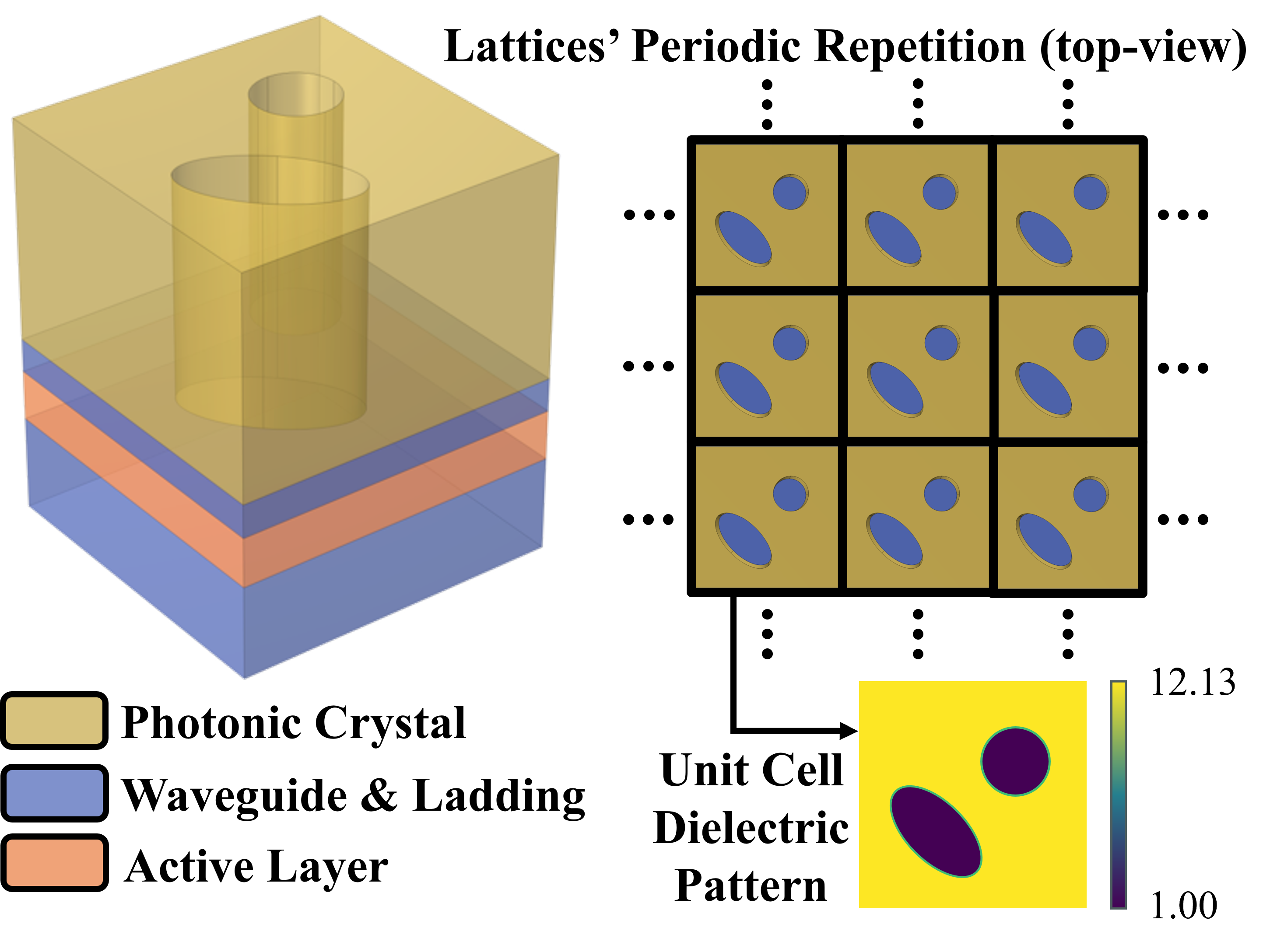}
    \captionsetup{belowskip=0pt}
    \caption{Schematic of the PhC lattice and its unit-cell representation. The unit-cell dielectric pattern is used as the model input, with colors indicating the dielectric constant.}
    \label{fig:pcsel_demo}
\end{figure}

Photonic-crystal surface-emitting lasers (PCSELs) \cite{hirose2014watt, noda2017photonic, yoshida2023high, noda2024photonic} have attracted substantial attention for their ability to simultaneously deliver high output power, excellent beam quality, and a compact device size. As illustrated in Figure~\ref{fig:pcsel_demo}, a PCSEL employs a two-dimensional photonic-crystal (PhC) layer formed by periodically repeated unit-cell geometries, whose dielectric modulation shapes the resonant modes and serves as a primary determinant of PCSELs' physical properties \cite{imada1999coherent, miyai2006lasers, kurosaka2010chip, noda2023high, noda2024photonic}. Therefore, efficient exploration of the vast PhC lattice design space is central to PCSEL development. Accordingly, we address this problem by predicting target physical properties directly from the unit-cell dielectric patterns of PhC lattices.

Efficient exploration of this design space faces two major challenges. First, physics-based simulations \cite{yokoyama2005finite, song2018first}, including the relatively efficient coupled-wave theory (CWT) \cite{liang2011three, liang2012three, liang2013three, inoue2019comprehensive, katsuno2021self}, require solving lattice-specific partial differential equations iteratively, making large-scale virtual screening prohibitively expensive. Second, the mapping from unit cells to physical properties is highly structured by underlying physics rather than governed by the visual similarities of lattices alone. CWT indicates that the resonant mode formation and out-of-plane radiation in PCSELs are largely governed by the spectral components and asymmetric structural components of the PhC unit-cell dielectric patterns, respectively \cite{liang2011three, liang2012three, yoshida2019double, wang2022symmetry, inoue2022general}. As a result, small geometric perturbations can induce sharp property variations in certain regions of the design space. Therefore, a practical AI surrogate needs not only to accelerate prediction, but also to represent the spectral and asymmetric structural information underlying such locally structure-sensitive responses.

Unfortunately, existing AI surrogates underexploit these physics-aligned structures, leading to a representation mismatch for PCSEL property prediction. Vision-based models, such as POST \cite{xin2025post}, learn from spatial dielectric patterns but lack an explicit spectral representation of the unit-cell geometries. Conversely, frequency-based models \cite{huang2026towards} introduce spectral information, yet their spectral features are not designed to preserve the phase covariance induced by random lattice shifts, making their spectral features difficult to exploit consistently. Moreover, although asymmetric structures dictate out-of-plane radiation in PCSELs, existing methods do not explicitly organize symmetry-resolved structural information. These limitations severely restrict their ability to learn accurate and reliable lattice-to-property mappings, especially in structure-sensitive regions where visually similar unit cells may exhibit substantially different physical behaviors.

To address these challenges, we propose the \textbf{D}ual-\textbf{D}omain \textbf{S}ymmetry-Aware \textbf{Net}work (\textbf{DDSNet}) for PCSEL property prediction. We translate the above CWT-indicated requirements into two physics-aligned inductive biases integrated into the network architecture. First, DDSNet introduces translation-equivariant spectral filtering as a frequency-domain inductive bias, enabling the network to exploit spectral components of PhC lattices while respecting the invariance of PCSEL properties under periodic lattice translations. Second, it imposes a symmetry-induced structural prior by separating PhC lattices' dielectric patterns into irreducible representation (irrep)-associated components and processing these components sperately. This design enables DDSNet to learn the property-specific contributions of different symmetry-resolved structural components. Together, these dual-domain mechanisms allow DDSNet to capture physically meaningful structure-property relationships and significantly improve its predictive performance and reliability in structure-sensitive regions.

Our main contributions are summarized as follows:
\begin{itemize}
    \item We propose DDSNet, a dual-domain symmetry-aware surrogate for PCSEL property prediction, which seamlessly integrates translation-equivariant spectral filtering and symmetry-induced structural priors to translate physical constraints into deep learning architectures.
    \item Through component masking analyses, we demonstrate that DDSNet successfully learns the property-specific dependencies of irrep-associated components, showing strict alignment with theoretical physical priors.
    \item DDSNet achieves a \(10^5\times\) acceleration over numerical CWT solvers while consistently outperforming existing AI baselines in property prediction and high-throughput screening (HTS), demonstrating superior stability in structure-sensitive regions.
\end{itemize}

\section{Related Work}

\noindent\textbf{Learning-based Modeling in Photonics}
Learning-based methods have emerged as powerful tools for modeling photonic devices, significantly reducing the reliance on computationally expensive simulations. Early studies primarily employed multilayer perceptrons (MLPs) to map structural parameters to optical responses \cite{peurifoy2018nanophotonic,malkiel2018plasmonic,an2019objective}. As photonic structures were represented by dielectric patterns, vision models were introduced to extract their spatial features \cite{jiang2021deep,ma2021deep,wiecha2021deep}. Recent advancements have further integrated neural operators and physics-informed networks to predict electromagnetic field with physical consistency \cite{gu2022neurolight,augenstein2023neural,ma2025pic2osim,lynch2025physics}. Beyond forward prediction, data-driven inverse design frameworks, ranging from tandem networks to advanced generative models, have been extensively developed to generate structures satisfying target responses \cite{liu2018training,tahersima2018deep,liu2018generative,yeung2020global,ma2022benchmark,seo2026physicsguided,tsukerman2025diffusion,marzban2026representation}.

\noindent\textbf{AI-assisted PCSEL Design}
Despite broad advances in general photonics, AI-assisted research tailored for PCSELs remains limited. For property prediction, \cite{huang2026towards} explored CWT-based data-driven modeling. POST \cite{xin2025post} established the current state-of-the-art by formulating PhC lattices as generic images and processing them via a Swin Transformer backbone \cite{DBLP:conf/iccv/LiuL00W0LG21}. For design exploration, sequential decision-making \cite{zhang2024pit} and LLM-assisted collaborative workflows \cite{li2023llm} have been proposed. 

As discussed in Section \ref{introduction}, existing PCSEL property prediction approaches underexploit critical physical priors, lacking either spectral inductive biases or translation equivariance and failing to preserve asymmetric structural information. To bridge this gap, DDSNet explicitly translates these CWT-derived priors into inductive bias and architectural designs, enabling more reliable learning of PCSEL structure-property relationships.

\section{Methodology}
\label{method}

\begin{table*}[t]
\centering
\small
\renewcommand{\arraystretch}{1.15} 
\caption{Irreducible representations of the C4v point group and their decomposition formulae.}
\label{tab:c4v_irreps}
\begin{tabular}{c l l}
\toprule[1.2pt]
\textbf{Irrep} & \textbf{Structural Interpretation} & \textbf{Symmetry Decomposition Formula (Projection Operator $P^\mu X$)} \\
\midrule

$A_1$ & Isotropic Base (Fully Symmetric)
& $X_{A_1} = (C_4^0 + C_4^1 + C_4^2 + C_4^3 + \sigma_x + \sigma_y + \sigma_d + \sigma_{d'}) X / 8$ \\

$A_2$ & Chiral / Rotational Symmetric
& $X_{A_2} = (C_4^0 + C_4^1 + C_4^2 + C_4^3 - \sigma_x - \sigma_y - \sigma_d - \sigma_{d'}) X / 8$ \\

$B_1$ & Inter-Axial Asymmetric
& $X_{B_1} = (C_4^0 - C_4^1 + C_4^2 - C_4^3 + \sigma_x + \sigma_y - \sigma_d - \sigma_{d'}) X / 8$ \\

$B_2$ & Inter-Diagonal Asymmetric
& $X_{B_2} = (C_4^0 - C_4^1 + C_4^2 - C_4^3 - \sigma_x - \sigma_y + \sigma_d + \sigma_{d'}) X / 8$ \\

$E$   & Spatial Gradient (2D)
& $X_{E_x} = (C_4^0 - C_4^2 + \sigma_x - \sigma_y) X / 4, \quad X_{E_y} = (C_4^0 - C_4^2 - \sigma_x + \sigma_y) X / 4$ \\

\bottomrule[1.2pt]
\multicolumn{3}{l}{\footnotesize $X$ represents the original input tensor.}
\end{tabular}
\end{table*}

\subsection{Preliminaries}
\noindent \textbf{General Discrete Symmetry Groups}
Discrete symmetry groups describe finite sets of geometric operations, such as rotations by fixed angles, reflections about prescribed axes, and their compositions. Given a group $G$, its regular representation assigns one group-indexed channel block to each group element, where group actions correspond to permutations of these channels. More generally, by complete reducibility of finite-group representations, any finite-dimensional representation space $V$ can be decomposed into a direct sum of irrep subspaces \cite{fulton2013representation}:
\begin{equation}
V \simeq
\bigoplus_{\lambda\in\Lambda}
\bigoplus_{j=1}^{m_\lambda}
V_{\lambda,j},
\quad
V_{\lambda,j}\cong V_\lambda ,
\end{equation}
where $\Lambda$ is the irrep set, $V_\lambda$ denotes the representation space of irrep $\lambda$, and $m_\lambda$ is its multiplicity. Each irrep follows a fixed transformation law under $G$, making irreps useful for organizing asymmetric structural information.

\noindent \textbf{The $p4m$ Symmetry and Irreducible Representations}
This work focuses on square-lattice PCSELs, whose unit-cell geometry is associated with the $p4m$ wallpaper symmetry. For the finite point-group component associated with this symmetry, we consider the $C_{4v}$ group:
\begin{equation}
C_{4v}
=
\{C_4^0,C_4^1,C_4^2,C_4^3,\sigma_x,\sigma_y,\sigma_d,\sigma_{d'}\}.
\end{equation}
Here, $C_4^k$ denotes $k$ successive $90^\circ$ rotations, and $\sigma_{(\cdot)}$ denotes reflection about the corresponding axis or diagonal. The regular representation of $C_{4v}$ can be understood as an enumeration of these eight discrete operations. For the $C_{4v}$ point group associated with $p4m$, we use the irrep set $\Lambda=\{A_1,A_2,B_1,B_2,E\}$ \cite{conway2016symmetries}. Table~\ref{tab:c4v_irreps} summarizes their structural interpretations and decomposition formulae. For a feature space carrying a $C_{4v}$ representation, a sample $X$ can be projected onto symmetry-adapted components:
\begin{equation}
X =
X_{A_1}+X_{A_2}+X_{B_1}+X_{B_2}+X_{E_x}+X_{E_y},
\end{equation}
where $X_\lambda$ denotes the component associated with irrep $\lambda$. An example projection of a lattice onto these irrep components is provided in Appendix \ref{appendix:irrep_decomposition}.

\noindent \textbf{Invariance of scalar PCSEL properties} In addition to point-group operations, a periodic unit-cell representation admits equivalent periodic translations. Such translations change the coordinate origin of the unit cell but preserve the underlying periodic lattice. For scalar PCSEL properties, equivalent periodic translations and $C_{4v}$ operations leave the target unchanged~\cite{liang2011three,xin2025post}:
\begin{equation}
y(s\cdot X)=y(X),
\quad
\forall s\in\mathcal{T}\cup C_{4v},
\end{equation}
where $\mathcal{T}$ denotes the set of periodic translations. Therefore, the final predictor should be invariant to these equivalent transformations. Intermediate features, however, may transform equivariantly according to prescribed representations, allowing the network to retain symmetry-resolved structural information before the final regression.

\begin{figure*}
    \centering
    \includegraphics[width=0.9\textwidth]{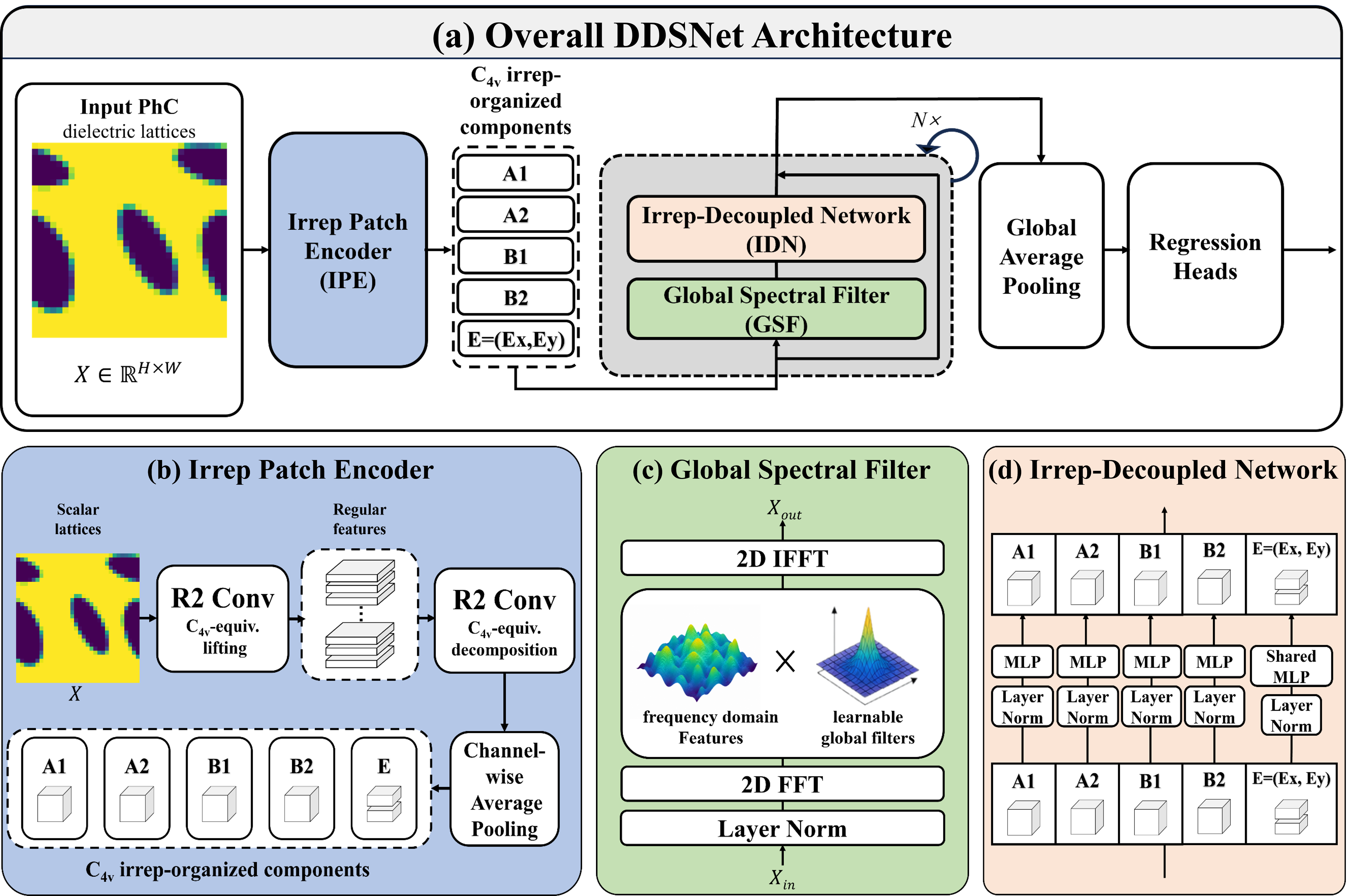}
    \caption{The framework of DDSNet.}
    \label{fig:overall_framework}
\end{figure*}

\subsection{Overall Framework}
\label{overall_framework}

Figure~\ref{fig:overall_framework} illustrates the overall architecture of DDSNet. Given a PhC unit-cell dielectric pattern, DDSNet predicts scalar physical properties through a dual-domain, symmetry-aware backbone. The Irrep Patch Encoder (IPE) first maps the lattice into irrep-organized token groups. The resulting tokens are then processed by stacked DDSNet blocks, each containing a Global Spectral Filter (GSF) and an Irrep-Decoupled Network (IDN). Finally, global average pooling and lightweight regression heads map the aggregated features to target properties.

\subsection{Translation-Equivariant Spectral Filtering}
\label{spectral_bias}

\noindent \textbf{Spectral motivation and translation constraint} CWT describes PCSELs by expanding the lattice dielectric distribution and optical modes into spectral components, reformulating wave-medium interaction as frequency-domain momentum matching~\cite{liang2011three,liang2012three,katsuno2021self}. This formulation suggests that spectral responses of the unit-cell dielectric pattern provide a physics-aligned representation for PCSEL property prediction, since coupling-related structural factors are more directly exposed in the spectral domain than in raw spatial domain, especially when visually similar lattices differ in physically important spectral components. Therefore, DDSNet injects spectral inductive bias into the vision backbone through spectral filtering.

However, spectral modeling must also respect the periodic nature of unit-cell representations. A periodic translation only changes the coordinate origin of the unit cell and doesn't alter PCSELs' scalar properties. According to the Fourier shift theorem, such a translation induces frequency-dependent phase modulation in the spectral domain:
\begin{equation}
\mathcal{F}\!\left(T_{\boldsymbol{\tau}}X\right)[\mathbf{k}]
=
\exp\!\left(-i\,\mathbf{k}\cdot\boldsymbol{\tau}\right)
\mathcal{F}(X)[\mathbf{k}],
\end{equation} 
where $T_{\boldsymbol{\tau}}$ denotes a periodic translation by $\boldsymbol{\tau}$, and $\mathbf{k}$ denotes a frequency index. Thus, a spectral module should modulate spectral components without breaking this phase covariance. Otherwise, the model has to learn to compensate for artificial phase inconsistencies among equivalent lattices. This motivates translation-equivariant spectral filtering.

\noindent \textbf{Global Spectral Filter} Inspired by the frequency filtering mechanism of GFNet~\cite{DBLP:conf/nips/RaoZZLZ21}, DDSNet introduces a Global Spectral Filter (GSF) to implement spectral inductive bias. For an input latent feature $X_{{in}}\in\mathbb{R}^{H\times W\times d}$, a GSF layer is formulated as
\begin{equation}
X_{{out}}
=
\Phi_{{GSF}}(X_{{in}})
=
\mathcal{F}^{-1}
\left(
K \odot \mathcal{F}(X_{{in}})
\right),
\end{equation}
where $\mathcal{F}$ and $\mathcal{F}^{-1}$ denote the 2D Fast Fourier Transform (FFT) and its inverse, respectively, $\odot$ denotes element-wise multiplication, and $K\in\mathbb{C}^{H\times W\times d}$ is a learnable complex-valued filter. The filter $K$ assigns frequency-specific complex weights to latent features, enabling the model to adaptively modulate spectral components relevant to PCSEL property prediction. A key property of this element-wise filtering is translation equivariance. We formalize it as follows:

\noindent\textbf{Proposition 1.}
\emph{Element-wise multiplication in the frequency domain is mathematically equivalent to a global circular convolution in the spatial domain, achieving equivariance with respect to circular spatial translations:}
\begin{equation}
\Phi_{{GSF}}(T_{\Delta\mathbf{x}}X)
=
T_{\Delta\mathbf{x}}\Phi_{{GSF}}(X).
\end{equation}

\noindent\emph{Proof.} See Appendix~\ref{appendix:PP1}.

For PCSEL lattices, periodic translations preserve the intrinsic structure-property relationship. Since GSF performs frequency-wise modulation, a spatial shift of the input induces the corresponding shift of the output feature map, rather than uncontrolled phase changes across spectral components. This property allows DDSNet to exploit spectral information while maintaining translation-consistent representations under lattice translations.

\noindent \textbf{Comparison with MLP-based spectral filtering} The above equivariance property also explains why not all dual-domain spectral modules are equally suitable for PCSEL property prediction. MLP-based spectral modules, such as AFNO~\cite{DBLP:conf/iclr/GuibasMLTAC22} and AFFNet~\cite{DBLP:conf/iccv/00140LZLG23}, apply nonlinear transformations to Fourier coefficients. Although such dense spectral mixing can enhance global token mixing, they may disrupt the phase covariance induced by lattice shifts, overlook the frequency-specific contributions suggested by CWT, and mix irrep-organized components before branch-wise nonlinear processing. These considerations motivate our usage of element-wise spectral filtering as the spectral module in DDSNet.

\subsection{Symmetry-Induced Structural Prior}
\label{asym_decomp}

\noindent \textbf{Motivation for symmetry-induced structural prior} CWT and previous PCSEL studies indicate that symmetry breaking structures strongly affect out of plane radiation. However, conventional analyses usually examine only few manually designed geometries through expensive simulations, making it difficult to systematically quantify how different asymmetric components contribute to target properties \cite{wang2022symmetry, inoue2022general}. DDSNet therefore introduces a symmetry-induced structural prior based on $C_{4v}$ irreps, which organizes lattice features into irrep-organized components and allows the network to learn property-specific dependencies across symmetry modes. Furthermore, strict $C_{4v}$ equivariance is enforced only in the shallow Irrep Patch Encoder (IPE), where these components are constructed. Subsequent feature-extraction blocks relax this constraint and process different components through separate branches to retain  representation capacity.

\noindent \textbf{Irrep Patch Encoder} To construct irrep-organized components, Irrep Patch Encoder (IPE) uses steerable convolutions~\cite{DBLP:conf/iclr/CohenW17}. Given a symmetry group $G$, the input and output feature channels are assigned representation types $\rho_{{in}}$ and $\rho_{{out}}$, which specify how these channels transform under each operation $g\in G$. For a convolutional layer to be equivariant from input type $\rho_{{in}}$ to output type $\rho_{{out}}$, its kernel $\kappa$ must satisfy
\begin{equation}
\kappa(g\mathbf{x})
=
\rho_{{out}}(g)\kappa(\mathbf{x})\rho_{{in}}(g)^{-1},
\quad \forall g\in G .
\end{equation}
This constraint defines the admissible kernel space for the chosen representation pair.
Steerable CNNs construct a basis of this space and parameterize the learnable kernel as
\begin{equation}
\kappa(\mathbf{x})
=
\sum_i \theta_i B_i(\mathbf{x}),
\end{equation}
where $B_i$ are basis kernels determined by the equivariance constraint, and $\theta_i$ are learnable coefficients.
Thus, changing the input-output representation pair changes both the admissible kernel basis and the transformation behavior of the output features.

IPE contains three steerable convolutional layers, whose structural representation flow is
\begin{equation}
\rho_{{triv}}
\;\longrightarrow\;
\rho_{{reg}}
\;\longrightarrow\;
\rho_{{reg}}
\;\longrightarrow\;
\rho_{{irrep}},
\end{equation}
where $\rho_{{triv}}$ is the trivial representation of the scalar dielectric input, $\rho_{{reg}}$ is the regular representation of $C_{4v}$, and $\rho_{{irrep}}$ denotes the irrep-organized representation associated with $A_1$, $A_2$, $B_1$, $B_2$, and $E$.

\noindent (1) \textbf{Lifting and processing in regular representation} The first layer lifts the scalar lattice input into regular features. Given a base kernel $\kappa$, IPE generates a set of symmetry-related filters by applying all operations in $C_{4v}$:
\begin{equation}
\kappa_g(\mathbf{r})
=
\kappa(g^{-1}\mathbf{r}),
\quad g\in C_{4v},
\end{equation}
where $\mathbf{r}$ denotes the relative coordinate inside the convolutional kernel. Applying these transformed filters to the input produces response channels indexed by the eight operations of $C_{4v}$. In other words, if the feature dimension is $d$, the regular feature channels can be viewed as $d/8$ groups, and each group stores responses to the same local pattern under eight rotated or reflected views. When the input is transformed by a $C_{4v}$ operation, these eight response channels are re-ordered according to the same group action, rather than being mixed arbitrarily. This is the regular representation used in IPE: it stores symmetry-related local responses in an explicitly group-indexed channel structure.

The second layer maps regular features to regular features. It further refines local patterns while preserving the same group-indexed channel organization. Since group actions in the regular representation correspond to channel permutations, standard point-wise nonlinearities can be applied without destroying the equivariant feature organization.

\noindent (2) \textbf{Mapping to irrep-organized components} The final layer converts regular features into irrep-organized components, whose
feature groups are associated with $C_{4v}$ irreps. Since each feature group in regular representation contains eight transformed views induced by the operations of $C_{4v}$, the final steerable convolution learns canonical local filters and generates their transformed copies across the eight views, while combining the resulting responses according to the symmetry pattern of the target irrep. For example, components with symmetric behavior aggregate symmetry-related responses, while components with antisymmetric behavior contrast them through substraction. This mechanism can be understood as a learned convolutional version of the projection rules in Table~\ref{tab:c4v_irreps}: the table gives fixed global combinations of transformed inputs, whereas IPE learns local filters whose responses are combined into the same types of symmetry-resolved components. Therefore, the final IPE layer produces irrep-organized feature groups rather than arbitrary dense mixtures of regular channels. The produced components are then passed to irrep-decoupled blocks for branch-wise nonlinear processing.

\noindent \textbf{Irrep-Decoupled Network} After IPE constructs irrep-organized components, Irrep-Decoupled Network (IDN) prevents them from being re-entangled by dense channel mixing. It uses five parallel MLP branches to process features originating from $A_1$, $A_2$, $B_1$, $B_2$, and $E$, respectively. For the two-dimensional $E$ representation, the $E_x$ and $E_y$ subchannels share the same MLP parameters, reflecting the equivalence of the two in-plane axes of the square lattice. The processed branches are then combined for scalar property prediction. This design preserves the symmetry-induced structural prior while allowing nonlinear interactions to be learned after branch-wise processing.

\section{Experiments}

\subsection{Evaluation Setup}
\label{eval_setup}

\textbf{Datasets and Metrics} We evaluate DDSNet on the open-source PCSEL dataset provided by POST~\cite{xin2025post}, which contains lattice-property pairs generated by CWT simulations of square-lattice PCSELs. Each sample is a $32 \times 32$ real-valued matrix representing the dielectric pattern of PhC unit cell. Following POST, we predict two scalar physical properties: surface-emitting efficiency ($SEE$), which measures effective surface-output conversion, and $\log Q$, where $Q$ characterizes optical confinement and is log-transformed due to its large dynamic range.  We report the coefficient of determination ($R^2$) for both $SEE$ and $\log Q$. Then, to evaluate candidate-ranking consistency for HTS, we also report Spearman's rank correlation coefficient ($\rho$) on the composite score (CS), which equally combines standardized $SEE$ and $\log Q$. Detailed definitions of $SEE$, $Q$ and $CS$ are provided in Appendix~\ref{appendix:metrics}.

We compare DDSNet with three categories of baselines. Frequency-based methods~\cite{huang2026towards} process spectral inputs. Spatial baselines treat dielectric patterns as images, including CNNs, Steerable CNNs~\cite{DBLP:conf/iclr/CohenW17}, ViTs~\cite{DBLP:conf/icml/TouvronCDMSJ21,DBLP:conf/icml/dAscoliTLMBS21,DBLP:conf/iccv/TouvronCSSJ21}, and POST~\cite{xin2025post}, the previous strongest PCSEL-specific surrogate. We further include dual-domain baselines, including AFFNet, GFNet, and SpectFormer~\cite{DBLP:conf/iccv/00140LZLG23,DBLP:conf/nips/RaoZZLZ21,DBLP:conf/wacv/PatroNA25}, to evaluate the effect of spectral inductive bias and provide direct comparisons for DDSNet's symmetry-resolved feature organization.

\label{main_result}
\begin{table*}[t]
\centering
\caption{PCSEL Property Prediction Experiment}
\label{tab:main_result}
\begin{tabular}{ll c p{2.6cm}<{\centering} p{2.6cm}<{\centering} p{2.8cm}<{\centering}}
\toprule[1pt] 
\textbf{Category} & \textbf{Model} & \textbf{Params (M)} & \textbf{$R^2$ of $SEE$ (std)} & \textbf{$R^2$ of logQ (std)} & \textbf{Spearman $\rho$ (std)}\\
\midrule
Freq.-based   & Huang et al.* & - & 0.595                    & 0.784                  & -  \\
\midrule
\multirow{7}{*}{\begin{tabular}[c]{@{}l@{}}Spatial-\\based\end{tabular}} 
                  & AlexNet* & - & 0.667                    & 0.817             &  -    \\
                  & DeiT-Ti* & - & 0.749                    & 0.869              &  -   \\
                  & ConViT-Ti* & - & 0.731                    & 0.882              &  -   \\
                  & CaiT-S24* & - & 0.763                    & 0.883              &  -   \\
                  & ResNet          & 11.218 & 0.7375 (0.0153)         & 0.8659 (0.0117)      &   0.8059 (0.0093) \\
                  & Steerable CNN   & \underline{0.383} & 0.7096 (0.0168)         & 0.8810 (0.0049)      &   0.7999 (0.0101) \\
                  & POST            & 5.116 & 0.7905 (0.0104)         & 0.9024 (0.0086)      &   0.8610 (0.0140) \\
\midrule
\multirow{4}{*}{\begin{tabular}[c]{@{}l@{}}Dual-\\Domain\end{tabular}} 
                  & AFFNet          & 1.473 & 0.7627 (0.0086)         & 0.8777 (0.0054)    &    0.8336 (0.0055)  \\
                  & GFNet           & 0.896 & \underline{0.8526 (0.0179)} & \underline{0.9410 (0.0089)} & \underline{0.9128 (0.0089)}  \\
                  & SpectFormer     & 1.119 & 0.8469 (0.0204) & 0.9346 (0.0046) &  0.9060 (0.0095) \\
                  & DDSNet          & \textbf{0.243} & \textbf{0.8835 (0.0078)} & \textbf{0.9527 (0.0042)} & \textbf{0.9363 (0.0030)}\\
\bottomrule[1pt] 
\multicolumn{6}{l}{\footnotesize * Results cited from original reports. "-" denotes metrics or parameter counts that were unavailable.}
\end{tabular}
\end{table*}

\subsection{Physical Property Prediction Experiment} 

The experiment results are presented in Table~\ref{tab:main_result}. DDSNet achieves the best performance across all metrics, outperforming POST, the strongest PCSEL-specific method, and extended dual-domain baselines. It also achieves the highest Spearman $\rho$ on CS, indicating stronger candidate-ranking consistency for the HTS task in Section~\ref{hts_result}. Based on these results, we highlight two observations:

\noindent \textbf{Benefit of spectral inductive bias}: Dual-domain baselines generally outperform spatial baselines, indicating the efficacy of spectral inductive biases for PCSEL property prediction. In particular, GFNet and SpectFormer show clear gains over POST, suggesting that translation-equivariant spectral filtering helps capture optical coupling patterns that are difficult to infer from spatial dielectric patterns alone.

\noindent \textbf{Impact of symmetry-resolved feature organization}: DDSNet further outperforms the dual-domain baselines, with a 3.6\% improvement in $R^2$ of $SEE$ over GFNet. Since DDSNet builds on the same element-wise spectral filtering principle as GFNet, this comparison provides evidence for the additional benefit of $C_{4v}$ irrep-associated feature organization. By organizing features into symmetry-resolved components and processing them through irrep-associated branches, DDSNet better captures symmetry-breaking structures related to PCSEL radiation. Meanwhile, its much smaller parameter count indicates that the improvement is driven by physics-aligned inductive biases.

For subsequent in-depth analyses, we focus on POST and GFNet as representatives from the spatial and dual-domain categories respectively, alongside DDSNet.

\begin{figure}[h]
    \centering
    \includegraphics[width=\linewidth]{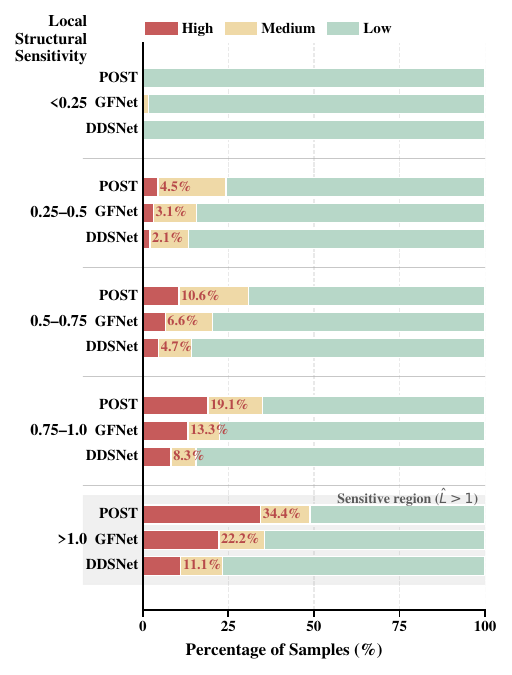}
    \caption{
    Error distribution across regions with different local structural sensitivities. 
    }
    \label{fig:diff_bin}
\end{figure}

\subsection{Investigation of Structure-Sensitive Regions}
\label{sample_bin}

We further test whether models can distinguish visually similar PhC lattices with distinct physical properties. To identify such cases, we define \emph{local structural sensitivity} $\hat{L}$, a metric that measures how sharply CS changes around a test sample relative to its local structural distances. Specifically, for each test sample, we retrieve visually similar training samples under a symmetry-invariant structural distance that accounts for periodic translations and $C_{4v}$ transformations. We then estimate $\hat{L}$ by averaging the top value of local empirical Lipschitz gradients, i.e., the ratios between CS differences and structural distances. Detailed definitions and implementations are provided in Appendix~\ref{appendix:structure_sensitive}. We partition test samples into sensitivity bins according to $\hat{L}$ and compare POST, GFNet, and DDSNet by their error-category distributions in each bin. Figure~\ref{fig:diff_bin} shows three observations:

\noindent \textbf{Stronger reliability across challenging regions}: DDSNet achieves the lowest high-error rates across almost all sensitivity levels. In the most sensitive regime ($>1$), its high-error rate is only 11.1\%, comparable to POST's performance in a much easier interval, 10.6\% at the 0.50-0.75 level.

\noindent \textbf{Efficacy in capturing sharp property shifts}: Since CS is computed from z-scored $SEE$ and $\log Q$, and the local structural distances are statistically $>1$, samples in highly sensitive regions reflect cases where standard deviation-level property changes exhibit in similar samples. DDSNet maintains a low high-error rate in these bins, suggesting that it is less misled by visual similarity between PhC lattices.

\noindent \textbf{Validation on model design}: DDSNet further improves over GFNet on the basis of GFNet's enhanced performance over POST, indicating the benefit of translation-equivariant spectral filtering and symmetry-induced structural prior.

\subsection{Probing Physical Inductive Biases} 
\label{ablation_component}

\begin{figure}[h]
    \centering
    \includegraphics[width=1\linewidth]{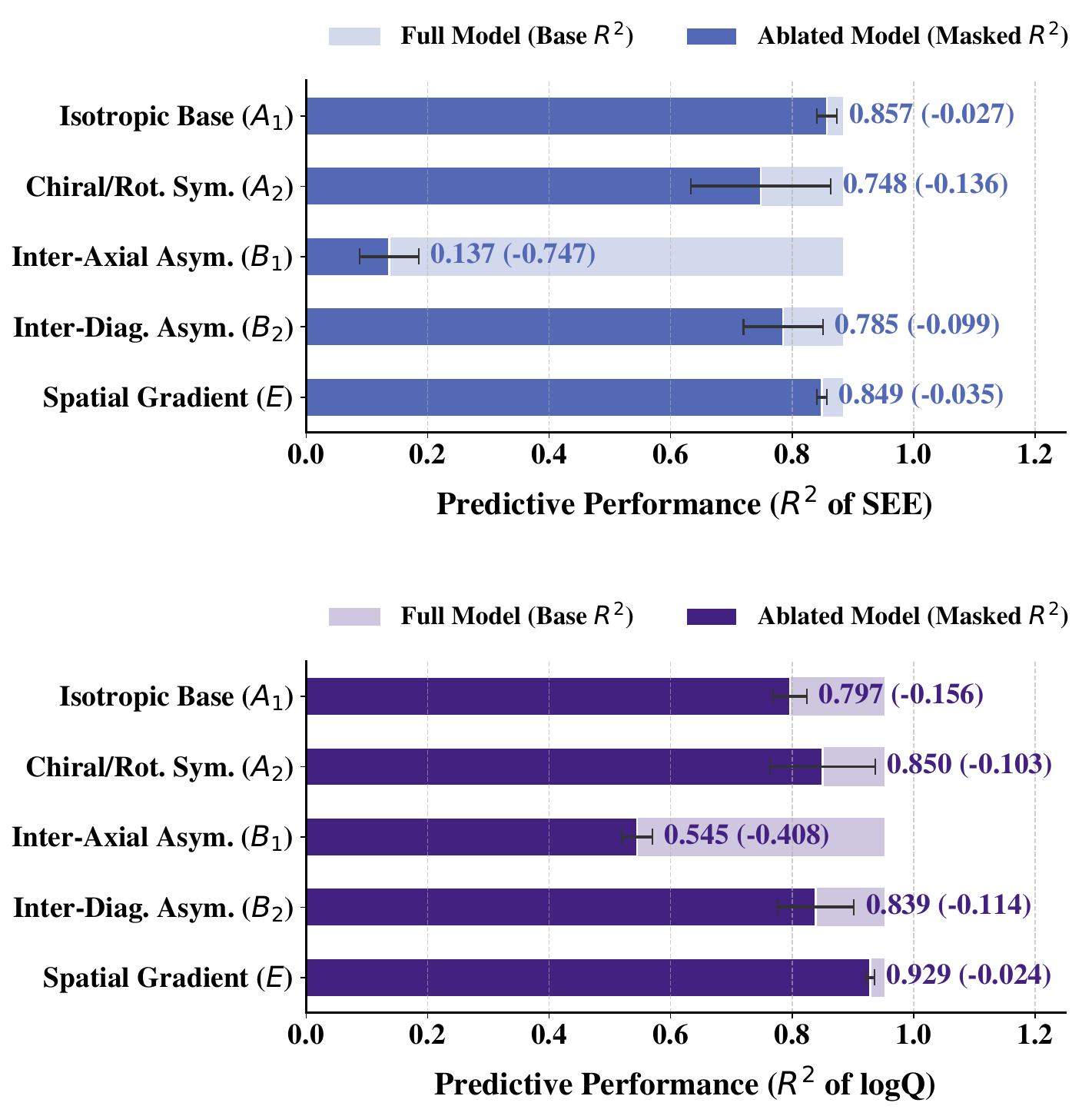}
        \caption{Physical property-specific dependencies on asymmetric components. By comparing the base performance (transparent bars) with the ablated performance (solid bars), we reveal the critical roles of specific irreps. The explicit performance drop demonstrates that different physical properties exhibit distinct dependencies on irreps, which are consistent with CWT-derived physical priors.}
    \label{fig:component}
\end{figure}

To verify whether DDSNet has learned physically meaningful representations, we conduct a component masking analysis. Because it organizes features according to irreps of the $C_{4v}$ group, for each irrep-associated component, we can mask its features before the final regression head and measure the resulting degradation in $R^2$. A larger degradation indicates that the corresponding symmetry component contributes more strongly to the predicted property. Figure~\ref{fig:component} shows that the learned component dependencies are consistent with CWT-based physical priors:

\noindent \textbf{Dominant role of the $B_1$ component}: Masking the $B_1$ component causes the largest performance degradation, indicating that DDSNet relies strongly on inter-axial asymmetry. This agrees with PCSEL physics: the dominant in-plane waves propagate along the orthogonal $x$- and $y$-directions, and B1 directly captures the imbalance between these principal axes. By contrast, $A_2$ and $B_2$ are odd with respect to the principal axes and weakly affect the primary Fourier components ($(K_x, 0)$ and $(0, K_y)$) involved in Bragg reflection and radiation coupling, supporting the model's heavy reliance on the $B_1$ component.

\noindent \textbf{Limited contribution of $A_1$ component to $SEE$}: Masking the fully symmetric $A_1$ component causes little degradation for SEE prediction. This is physically reasonable because $SEE$ is mainly governed by symmetry-breaking components that enable out-of-plane radiation, while the fully symmetric base alone does not encode the asymmetry required to break destructive interference.

\noindent \textbf{Weak effect of the $E$ component}: The $E$ component corresponds to a two-dimensional vector-like representation and captures directional, polarity-like variations, which can be intuitively viewed as translation-like centroid shifts of the unit-cell pattern. Masking it causes only minor performance degradation, suggesting that these components contribute weakly to the physical properties considered here. This is consistent with the invariance of $SEE$ and $Q$ under periodic translations and point-group operations.

\begin{figure}[h]
    \centering
    \begin{subfigure}{\linewidth}
        \centering
        \includegraphics[width=0.7\linewidth]{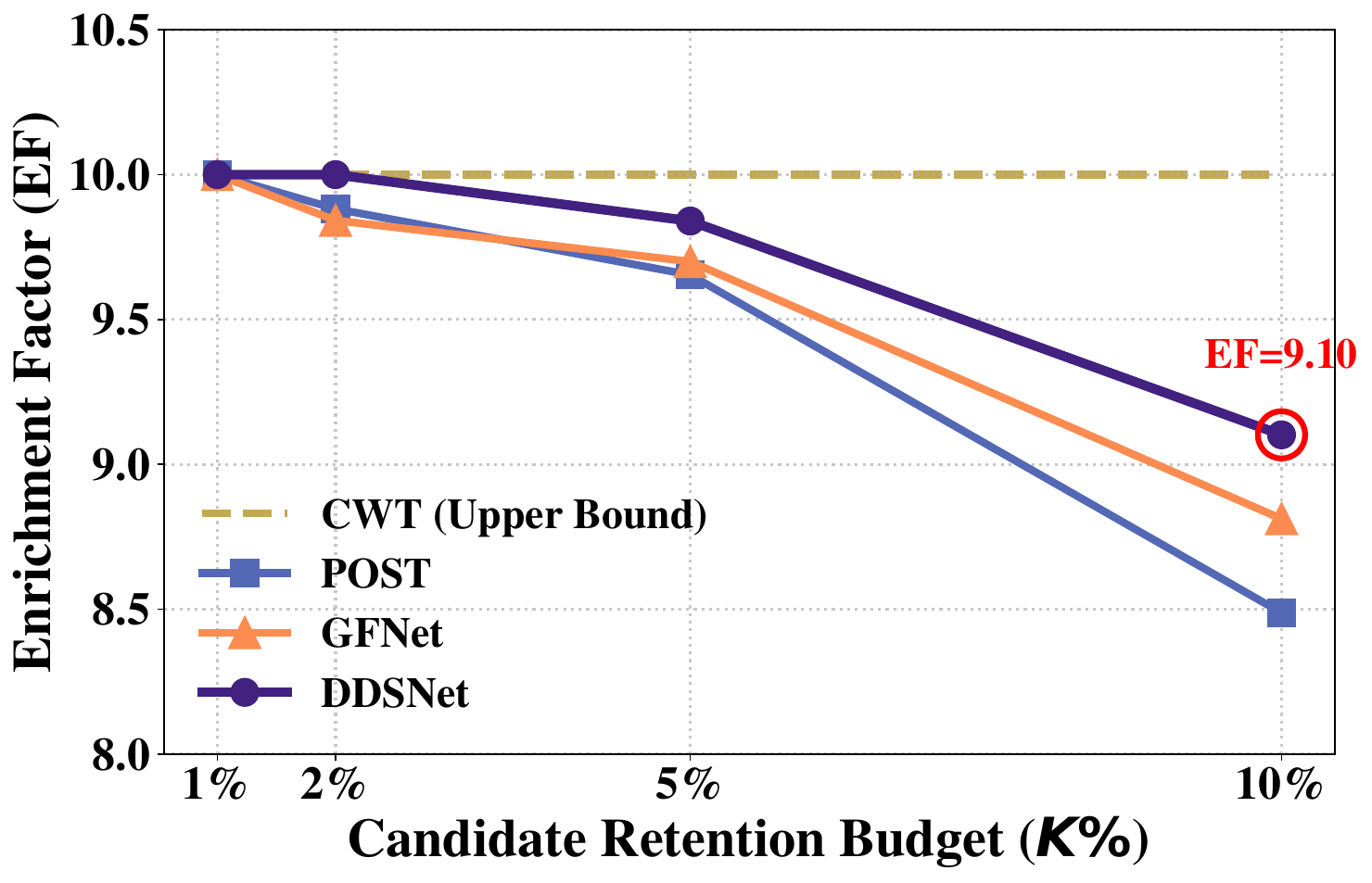}
        \caption{High-Throughput Screening Quality}
        \label{fig:hts_ef}
    \end{subfigure}
    
    \begin{subfigure}{\linewidth}
        \centering
        \includegraphics[width=0.7\linewidth]{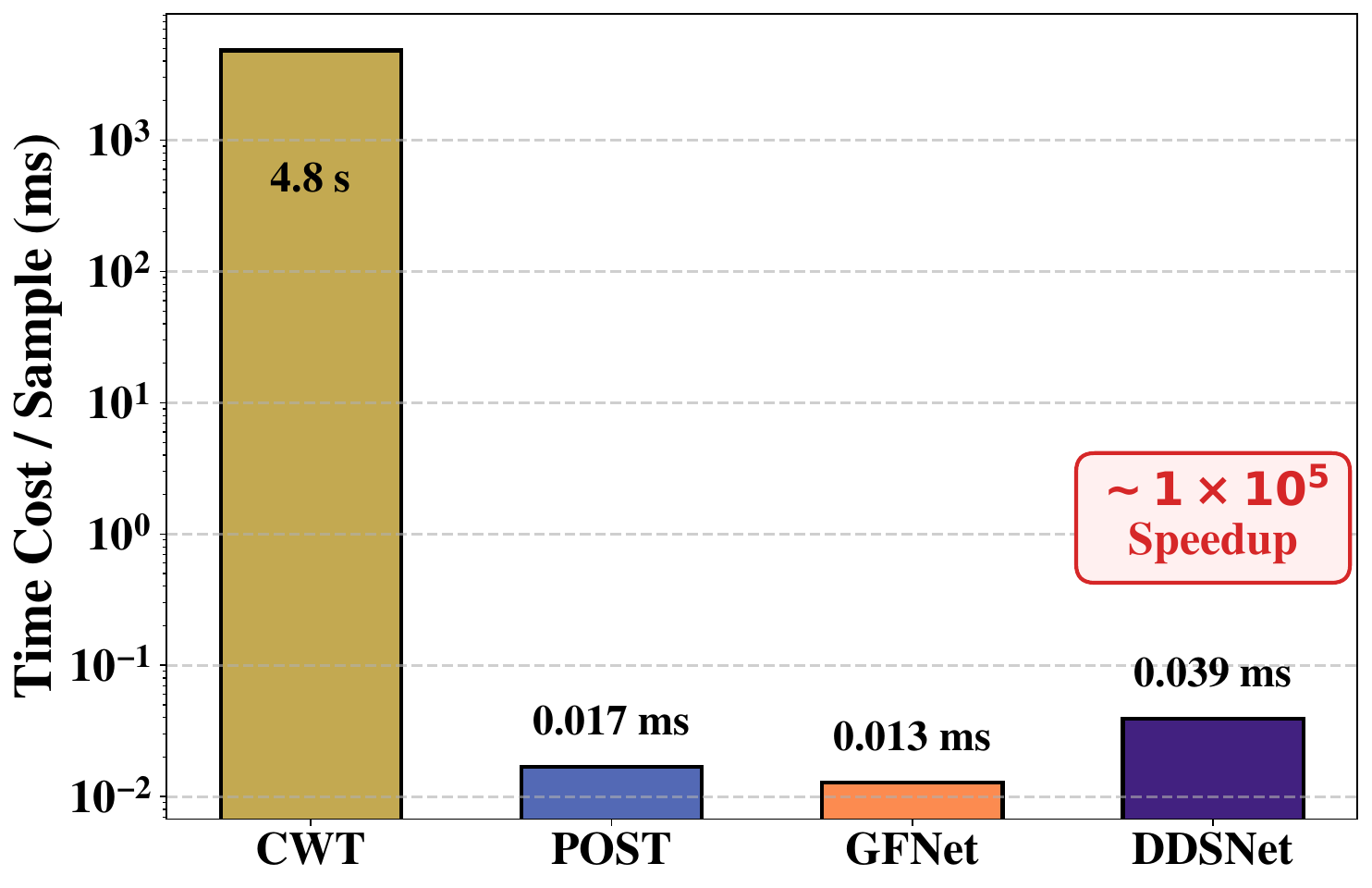}
        \caption{Inference Acceleration (Log Scale)}
        \label{fig:hts_latency}
    \end{subfigure}
    
    \caption{Performance evaluation in practical HTS scenarios. \textbf{(a)} DDSNet achieves the highest EF across various screening budgets, closely approaching the CWT upper bound. \textbf{(b)} Comparison of inference latency per sample. Compared to the CWT solver on a single CPU core, DDSNet on a single GPU demonstrates a $\sim 1.2 \times 10^5 $ acceleration.}
    \label{fig:hts_result}
\end{figure}

\subsection{Application: High-Throughput Screening}
\label{hts_result}
We further conduct a HTS experiment to simulate the practical usage of AI surrogates in PCSEL design. In HTS, candidate PhC structures are ranked by CS, and only the top $K\%$ candidates are retained for subsequent CWT simulation or device-level validation, where $K\%$ is termed the \emph{candidate retention budget}. We treat the top 10\% candidates ranked by ground-truth CS as true high-potential candidates, or true hits, and evaluate each surrogate by ranking test candidates using predicted CS. Following~\cite{truchon2007evaluating,wojcikowski2017performance}, the enrichment factor (EF) is utilized as main evaluation metric, which measures the fold enrichment of true hits over random selection. Detailed definitions and Implementations are provided in Appendix~\ref{appendix:hts}. The HTS results is reported in Figure~\ref{fig:hts_result}, leading to two observations:

\noindent \textbf{Massive computational acceleration}: All neural surrogates are substantially faster than numerical CWT evaluation. CWT requires about 4.8 s per candidate, whereas neural surrogates require only 0.013-0.039 ms under batched GPU inference. DDSNet takes 0.039 ms, yielding over $10^5\times$ speedup and enabling large-scale PhC lattice screening.

\noindent \textbf{Stronger candidate enrichment}: DDSNet achieves the best screening quality across candidate retention budgets. Since HTS serves as a pre-screening step, a moderately broad budget such as 10\% is practically important: it preserves room for additional requirements and fabrication tolerances. At this 10\% budget, DDSNet obtains the highest EF of 9.10, outperforming POST and GFNet, while approaching the upper bound of 10.0. Its consistently strong EF across 1\%-10\% budgets further indicates that DDSNet preserves high-potential PhC structures more reliably than the baselines.

\section{Conclusion}
We introduce DDSNet, a dual-domain symmetry-aware network for PCSEL physical property prediction in this paper. DDSNet integrates spectral inductive bias and symmetry-induced structural prior into a vision backbone, enabling the model to capture the wave-optics mechanisms that govern PCSEL structure-property relationships. Experiments show that DDSNet achieves over \(10^5\times\) acceleration over CWT simulations while significantly outperforming existing AI baselines in property prediction and high-throughput screening. Further analyses demonstrate that DDSNet is more robust in structure-sensitive regions and learns symmetry-component dependencies consistent with physical priors of CWT. These results suggest that translation-equivariant spectral filtering and symmetry-induced structural organization provide an effective route toward physically informed AI surrogates for large-scale PhC lattice screening.

\bibliography{aaai2027}

\newpage
\clearpage
\appendix

\twocolumn[%
    \begin{center}
    \fontsize{15}{18}\selectfont
    \textbf{DDSNet: Dual-Domain Symmetry-Aware Network for PCSEL Property Prediction (Supplementary Material)}\\
    \vspace{0.3cm}
    
    \fontsize{12}{16}\selectfont
    \textbf{
    Cen Chen\textsuperscript{\rm 1,\rm 2,*},
    Haitao Huang\textsuperscript{\rm 1,*},
    Jiazhi Mao\textsuperscript{\rm 3},
    Feifan Xu\textsuperscript{\rm 3},
    Zhe Zhuang\textsuperscript{\rm 3,\dag},
    Yuxiang Ren\textsuperscript{\rm 1,\dag}
    }\\
    \vspace{0.3cm}
    
    \textbf{\normalsize{\mdseries
    \textsuperscript{\rm 1}School of Intelligent Science and Technology, Nanjing University\\
    \textsuperscript{\rm 2}School of Information and Software Engineering, University of Electronic Science and Technology of China\\
    \textsuperscript{\rm 3}School of Integrated Circuits, Nanjing University\\
    ccen1613@gmail.com,
    casishaitaohuang@gmail.com,
    jzmao@smail.nju.edu.cn,
    ffxu@nju.edu.cn,
    zzhuang@nju.edu.cn,
    renyuxiang@nju.edu.cn
    }}\\
    \vspace{0.2cm}
    
    \normalsize{
    \textsuperscript{*}Equal contribution.
    \quad
    \textsuperscript{\dag}Corresponding authors.
    
    }
    
    \end{center}
    \vspace{0.5cm}
]

\section{Proof of Proposition 1}
\label{appendix:PP1}

For clarity, since GSF applies independently to all channels, we consider one feature channel. 
Let \(X\in\mathbb{C}^{H\times W}\), and denote its 2D discrete Fourier transform by
\begin{equation}
\widehat{X}[u,v]=\mathcal{F}(X)[u,v].
\end{equation}
For a periodic translation $T_{\Delta\mathbf{x}}$  with \(\Delta\mathbf{x}=(\Delta h,\Delta w)\), we have the shift property of the discrete Fourier transform,
\begin{equation}
\mathcal{F}(T_{\Delta\mathbf{x}}X)[u,v]
=
\exp\!\left(
-2\pi i\left(\frac{u\Delta h}{H}+\frac{v\Delta w}{W}\right)
\right)
\widehat{X}[u,v].
\end{equation}
Let
\begin{equation}
D_{\Delta\mathbf{x}}[u,v]
=
\exp\!\left(
-2\pi i\left(\frac{u\Delta h}{H}+\frac{v\Delta w}{W}\right)
\right).
\end{equation}
Then
\begin{equation}
\mathcal{F}(T_{\Delta\mathbf{x}}X)
=
D_{\Delta\mathbf{x}}\odot \mathcal{F}(X).
\end{equation}
Applying the GSF to the translated input gives
\begin{equation}
\begin{aligned}
\Phi_{{GSF}}(T_{\Delta\mathbf{x}}X)
&=
\mathcal{F}^{-1}
\left(
K\odot \mathcal{F}(T_{\Delta\mathbf{x}}X)
\right) \\
&=
\mathcal{F}^{-1}
\left(
K\odot D_{\Delta\mathbf{x}}\odot \mathcal{F}(X)
\right) \\
&=
\mathcal{F}^{-1}
\left(
D_{\Delta\mathbf{x}}\odot K\odot \mathcal{F}(X)
\right).
\end{aligned}
\end{equation}
Since \(K\) and \(D_{\Delta\mathbf{x}}\) are both applied element-wise in the frequency domain, they commute. Using the inverse shift property of the DFT, we obtain
\begin{equation}
\mathcal{F}^{-1}
\left(
D_{\Delta\mathbf{x}}\odot K\odot \mathcal{F}(X)
\right)
=
T_{\Delta\mathbf{x}}
\mathcal{F}^{-1}
\left(
K\odot \mathcal{F}(X)
\right).
\end{equation}
Therefore,
\begin{equation}
\Phi_{{GSF}}(T_{\Delta\mathbf{x}}X)
=
T_{\Delta\mathbf{x}}\Phi_{{GSF}}(X),
\end{equation}
which proves translation equivariance.

Moreover, by the convolution theorem, if
\begin{equation}
h=\mathcal{F}^{-1}(K),
\end{equation}
then
\begin{equation}
\Phi_{{GSF}}(X)
=
\mathcal{F}^{-1}
\left(
K\odot \mathcal{F}(X)
\right)
=
h *_{{circ}} X,
\end{equation}
where \(*_{{circ}}\) denotes circular convolution. Since \(K\) is applied independently to each feature channel, this operation is a depth-wise global circular convolution.

\begin{table}[t]
\centering
\small
\setlength{\tabcolsep}{4pt}
\renewcommand{\arraystretch}{1.08}
\caption{Main architecture and training hyperparameters.}
\label{tab:impl_hparams}
\begin{tabular}{l l}
\toprule
\textbf{Item} & \textbf{Setting} \\
\midrule
\multicolumn{2}{l}{\textit{Irrep Patch Encoder}} \\
Input size & $32\times 32$ \\
Patch size & $4\times 4$ \\
Token grid & $8\times 8$ \\
R2Conv layers & 3 \\
R2Conv kernel/stride & $3\times3$ / 1 \\
Padding & circular \\
Irrep channels & $[16,16,32,32,16]$ \\
Mappings & Triv.$\to$Reg., Reg.$\to$Reg., Reg.$\to$Irreps \\
\midrule
\multicolumn{2}{l}{\textit{DDSNet Backbone}} \\
Embedding dimension & 128 \\
Number of blocks & 8 \\
GSF filter size & $8\times5$ using rFFT/iFFT \\
MLP expansion ratio & 3 \\
DropPath rate & linearly increasing $0\to0.2$ \\
Regression head & one-hidden-layer MLP, hidden dim 32 \\
Task heads & shared backbone, separate heads \\
\midrule
\multicolumn{2}{l}{\textit{Optimization}} \\
Loss & $0.5\mathcal{L}_{SEE}+0.5\mathcal{L}_{\log Q}$ \\
Base loss & MSE \\
Optimizer & AdamW \\
Learning rate & $1\times10^{-4}$ \\
Weight decay & $1\times10^{-3}$ \\
Scheduler & cosine annealing \\
Epochs & 100 \\
\bottomrule
\end{tabular}
\end{table}

\section{Implementation Details}
We evaluate all models under the same five-fold protocol. For each fold, four folds are used for training, and the remaining fold is evenly split into validation and test sets. All baselines and DDSNet use the same random seeds, the same train/validation/test split and the same data augmentations in each fold. Reported standard deviations are computed over the five folds.

Because the original dataset is limited in scale, we apply label-preserving augmentation to the training split only. For each training lattice, we construct augmented variants using rotations or reflections from the $p4m$ group and periodic translations, resulting in a $36\times$ augmentation pool. Since the target scalar PCSEL properties are invariant to these transformations, the original labels are preserved. Validation and test samples are never augmented. The same augmented result is applied to all reproduced baselines and DDSNet for fair comparison.

Table~\ref{tab:impl_hparams} summarizes the main architecture and training hyperparameters of DDSNet. 
The IPE contains three R2Conv layers with circular padding. 
The first two layers lift and process regular features, while the last layer maps regular features into irrep-structured groups. 
After IPE, the $32\times32$ input is represented as an $8\times8$ token grid. 
DDSNet then stacks eight backbone blocks, each consisting of a Global Spectral Filter and an Irrep-Decoupled Network. 
For multi-task prediction, $SEE$ and $\log Q$ share the backbone but use separate lightweight regression heads.

\section{Implementation Details of Property Prediction and Ranking Experiments}
\label{appendix:metrics}

\noindent \textbf{Physical properties' definitions} Following POST~\cite{xin2025post}, we evaluate two CWT-computed scalar physical properties: surface-emitting efficiency ($SEE$) and quality factor ($Q$). The surface-emitting efficiency is defined as
\begin{equation}
SEE=\frac{P_{surface}}{P_{stim}},
\end{equation}
where $P_{surface}$ is the effective surface-emitted power and $P_{{stim}}$ is the stimulated emission power.
The quality factor is defined as
\begin{equation}
Q=\frac{\omega W}{P_{{stim}}},
\end{equation}
where $\omega$ is the resonant angular frequency and $W$ is the total energy stored in the cavity.
Since raw $Q$ values span several orders of magnitude, we predict $\log Q$ for stable optimization.

\noindent \textbf{Ranking-based evaluation setup} For ranking-based evaluation and HTS, we define a composite score (CS) by equally combining standardized $SEE$ and $\log Q$:
\begin{equation}
CS = 0.5 z_{SEE} + 0.5 z_{\log Q},
\end{equation}
where the $SEE$ and $\log Q$ values are standardized using the mean and std computed from the training split. The same training-split statistics are used to compute both the ground-truth CS from the CWT labels and the predicted CS from model outputs on the validation and test splits.

To evaluate candidate-ranking consistency, we sort test samples by their ground-truth CS and predicted CS, respectively. Then, we compute Spearman's rank correlation coefficient $\rho$ between these two rankings. This metric evaluates whether a model preserves the relative ordering of high-potential PCSEL candidates. The reported $\rho$ is averaged over five folds, with the std computed across five folds.

\section{Implementation Details of Structure-Sensitive Region Evaluation}
\label{appendix:structure_sensitive}

We evaluate model robustness in structure-sensitive regions by estimating the local empirical Lipschitz constant of the ground-truth structure-property mapping. 
The metric is computed in the original lattice space, since all evaluated models take the PhC lattice as input and predict the scalar physical properties as output. 
Therefore, a large empirical gradient indicates that a small structural perturbation can induce a large property shift, corresponding to an intrinsically difficult region for data-driven surrogate models.

\noindent \textbf{Local structural distance}
Because scalar PCSEL properties are invariant to point-group operations and periodic translations, we define the local structural distance between a test sample $x_i$ and a training sample $x_j$ as
\begin{equation}
d(x_i,x_j) = \sqrt{\min_{g\in C_{4v},\,\mathbf{v}} {MSE} \left(x_i,\, g(T_{\mathbf{v}}(x_j)) \right) },
\end{equation}
where $T_{\mathbf{v}}$ denotes periodic translation by vector $\mathbf{v}$ under periodic boundary conditions. 
The minimization over translations is efficiently computed using FFT-accelerated cross-correlation, and the minimization over $C_{4v}$ is performed by enumerating the eight point-group operations.

\noindent \textbf{Local structural sensitivity}
We use the difference between ground truth of CS to measure the property discrepancy. For the training sample $x_j$, which is visually similar to a test sample $x_i$, we compute the local Lipschitz empirical gradient, i.e. the ratio between the CS difference and the structural distance.
\begin{equation}
q_j
=
\frac{
|{CS}(x_i)-{CS}(x_j)|
}{
d(x_i,x_j)+\epsilon
},
\end{equation}
where $\epsilon$ is a small constant for numerical stability. 
To reduce sensitivity to local noise and computational overhead, the local structural sensitivity $\hat{L}_i$ of test sample $i$ is estimated by averaging the three largest empirical gradients among the 50 most similar training samples.

\noindent \textbf{Binning and error categorization} Test samples are partitioned according to $\hat{L}_i$ with a bin width of $0.25$. Due to sample sparsity in extremely sensitive regions, all samples with $\hat{L}_i>1$ are merged into a single bin. Within each bin, we evaluate each model by the absolute prediction error of CS:
\begin{equation}
e_i^{(m)}=\left|\widehat{CS}_i^{(m)}-CS_i\right|,
\end{equation}
where $CS_i$ is the ground-truth composite score and $\widehat{CS}_i^{(m)}$ is the composite score predicted by model $m$. Prediction errors are categorized into Low Error ($e_i^{(m)} \le 0.1$), Medium Error ($0.1<e_i^{(m)}<0.5$), and High Error ($e_i^{(m)} \ge 0.5$).

\section{Implementation Details of Probing Physical Inductive Biases}
\label{appendix:irrep_intervention}

We perform irrep intervention at inference time to quantify property-specific dependencies on asymmetric components. 
The intervention is applied after global average pooling and before the task-specific regression heads. 
For each irrep group $\lambda\in\{A_1,A_2,B_1,B_2,E\}$, we set all channels belonging to $\lambda$ to zero and keep the remaining features unchanged. 
For the two-dimensional irrep $E$, the two subchannels $E_x$ and $E_y$ are masked simultaneously. 
No model parameters are updated after masking.

For each masked model, we compute the $R^2$ score on the test set and report the performance degradation
\begin{equation}
\Delta R^2_{\lambda}
=
R^2_{{full}}
-
R^2_{{mask}(\lambda)}
\end{equation}
A larger $\Delta R^2_{\lambda}$ indicates that the corresponding irrep group contributes more strongly to the prediction. 
We compute $\Delta R^2_{\lambda}$ separately for $SEE$ and $\log Q$, and average the results over the five folds.

\section{Implementation Details of High-Throughput Screening}
\label{appendix:hts}

Following~\cite{truchon2007evaluating,wojcikowski2017performance}, we quantify screening performance by the enrichment factor:
\begin{equation}
EF_{K\%}=\frac{N^{K\%}_{{hits}}}{N^{{total}}_{{hits}}\times K\%},
\end{equation}
where $N^{K\%}_{{hits}}$ is the number of true hits retrieved within the selected top $K\%$ candidates, and $N^{{total}}_{{hits}}$ is the total number of true hits. An $EF$ of 1 corresponds to random selection, while a larger $EF$ indicates stronger enrichment of true high-potential structures. In each test fold, the top $10\%$ samples ranked by ground-truth CS are defined as true hits. For each surrogate, samples are ranked by predicted CS, and the top $K\%$ candidates are selected under candidate retention budgets $K\in\{1,2,5,10\}$.

For computational efficiency, we report the amortized evaluation time per candidate. The numerical CWT solver is evaluated using 30 processes under two AMD EPYC 9124 CPUs. And its runtime is reported as the wall-clock solver time per candidate. Neural surrogate models are evaluated on a single NVIDIA RTX 4090 GPU with batch size 256. For neural surrogates, we first run five warm-up passes and then average the runtime over 100 repeated evaluations on the test split, using CUDA synchronization before and after timing. The reported time per candidate is obtained by dividing the total inference time by the number of evaluated test samples.

\begin{figure*}[t]
    \centering
    \includegraphics[width=\textwidth]{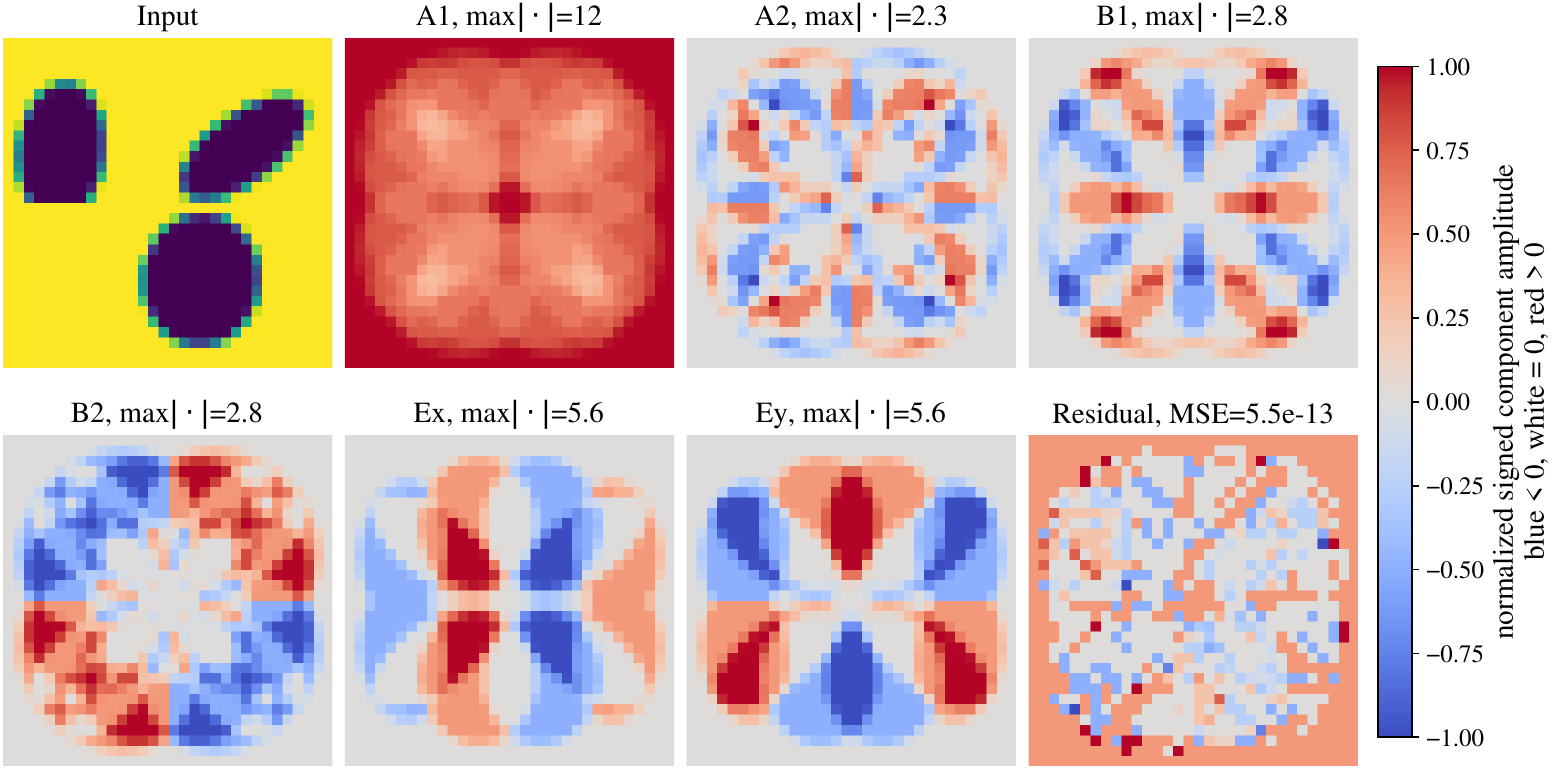}
    \caption{
    Example visualization of $C_{4v}$ irrep decomposition for a representative PhC lattice. 
    The input is decomposed into $A_1$, $A_2$, $B_1$, $B_2$, and $E=(E_x,E_y)$ components. 
    Components are normalized independently for visualization, and the residual indicates the reconstruction error after summing all components.
    }
    \label{fig:irrep_decomposition_example}
\end{figure*}

\section{Example of Irrep Decomposition}
\label{appendix:irrep_decomposition}

Figure~\ref{fig:irrep_decomposition_example} visualizes the decomposition of a representative PhC lattice into $C_{4v}$ irrep components. 
The input lattice is projected onto $A_1$, $A_2$, $B_1$, $B_2$, and the two coordinate channels of the two-dimensional irrep $E=(E_x,E_y)$. 
Each component is independently normalized for visualization, while the title reports its original maximum absolute amplitude. 
The residual map shows the reconstruction error after summing all projected components, confirming that the irrep components form a numerically accurate decomposition of the original lattice. 
This visualization is only used to illustrate the geometric meaning of different asymmetric components; the learned feature importance is analyzed separately through the intervention experiment in Section~\ref{ablation_component}.

\end{document}